# Co-evolution of point defects and Cr-rich nano-phase in binary Fe-20 at.% Cr alloy: A comprehensive investigation using positron annihilation spectroscopy and atom probe tomography

Sudip Kumar Sarkar [1*,3], Priya Maheshwari [2,3], P. K. Pujari [2,3], Aniruddha Biswas [1,3]

[1]Materials Science Division, [2]Radiochemistry Division, Bhabha Atomic Research Centre, Mumbai-40008, India

[3]Homi Bhabha National Institute, Mumbai-400094, India

[*]E-mail: s.sudip.iitg@gmail.com, sudips@barc.gov.in

**Abstract**

The role of point defects in temporal evolution of Cr-rich $\alpha'$ phase separation in binary Fe-20 at.% Cr alloy is elucidated by intercepting the long term (upto 1000 h at 773 K) aging at regular intervals and probing by a combination of atom probe tomography (APT) and positron annihilation spectroscopy (PAS). Since the Cr concentration of nano-scale $\alpha'$ phase in Fe-20 at.% Cr alloy increases continually on aging, the point defects play active role throughout the aging duration. The near-atomic resolution of APT and self-seeking ability of positrons towards point defects make this analysis possible. The difference of positron affinities between Fe and Cr enables identification of the Cr-rich nano-phases that are specifically associated with defects; they would otherwise be indiscernible in the absence of defect. Thus, the temporal evolution of Cr-rich nano-phase along with the associated point defects can be fully characterized at each stage of aging using APT and PAS, respectively. The combined APT-PAS study shows that Cr-rich $\alpha'$ precipitation is preceded by vacancy-Cr complexes that act as nucleation sites for Cr clusters, which in turn, cause an early rise in micro-hardness due to cluster hardening. Interestingly, this is accompanied by a significant rise in point defect concentration. Our results depict that the vacancies present in the core of the $\alpha'$ precipitates migrate towards the precipitates' interface during the course of aging and eventually get recovered resulting in virtually defect-free precipitates. This study chronicles the way the point defects shape the process of α′ phase separation throughout the entire length of aging.

## 1. Introduction

Solid state phase transformations and associated microstructural changes, especially the characteristics of the transformation products, often play key role in governing material properties [1–3]. A closer look also reveals the significant influence of point defects on many of these microstructural changes that are caused by phase transformations. It is understood that phase transformations involve diffusion of atoms, which in turn, are largely mediated through lattice defects, like vacancies. A large number of studies in the literature have detailed the role of point defects in explaining the age-hardening behaviour of different alloy systems, especially Fe- and Al-based alloys [4–12]. Point defects often act as preferred nucleating sites for pre-precipitate aggregation/clustering of solute atoms and thus, the kinetics of age-hardening can be controlled by optimizing the point defect concentration [12]. Not surprisingly, generation of excess vacancies facilitates solid state precipitation, as shown recently in a popular age-hardenable alloy system, Al-Cu [13]. The role of point defects is all the more obvious in case of radiation induced solute segregation, which is manifested by the accelerated kinetics of solute segregation and premature embrittlement of alloys under irradiation [9–11].

Besides age-hardening, phase separation is another important phase transformation process that involves potentially significant role of point defects and thus, merits detailed investigation. However, as discussed by Nie et al. in a seminal article, there are some notable differences between age-hardening and phase separation especially at the early stages of phase transformations that involve clustering of solutes [14]. The differences reportedly stem from the fact that, unlike the former, phase separation involves creation of a second phase having the same crystal structure and spatial orientation. As a result, in case of phase separation, the pre-precipitate solute clusters are expected to make seamless transition to nuclei of the precipitates [14,15]. Among the alloy systems that are commonly known to exhibit phase separation owing to the presence of miscibility gap in the phase diagram [16], Fe-Cr system [17] is particularly relevant because of its wide applicability, especially in nuclear industry in the form of high-Cr ferritic/martensitic (F/M) steels or oxide dispersion strengthened (ODS) ferritic steels [18–20]. These high-Cr alloys are among the candidate structural materials for Generation-IV nuclear reactors owing to their excellent corrosion resistance, low void swelling, low ductile to brittle transition temperature and good high temperature strength [18,19,21–25]. Under exposure of temperature and/or, radiation, the ferrite phase ($\alpha$) in these alloys undergoes phase separation into nano-scale ultrafine mixture of Cr-rich $\alpha'$ and Fe-rich $\alpha$ [26–28], which is known to follow either spinodal decomposition [29–31] or nucleation-growth mode [32–34] depending on the Cr-concentration of ferrite and temperature of operation [35]. This phase separation process results in hardening, loss of ductility and resultant embrittlement, and a convenient way to study $\alpha'$ phase separation is by using the binary Fe-Cr model alloys [32–34,36]. Binary model alloys help avoid the complexity that often arises due to the presence of other alloying elements. Expectedly, quite a few studies on $\alpha'$ precipitation in binary Fe-Cr alloys induced by thermal aging [32,37,38], and also by different types of radiations viz., neutron [33,34], proton [39], electron [40] are available in the literature. But, there has not been any systematic investigation on the role of point defects in $\alpha'$ phase separation in Fe-Cr system. Moreover, probing the role of point defects in $\alpha'$ phase separation in Fe-Cr system becomes especially pertinent because of the non-classical nature of nucleation-growth operative in this system [32]. The Cr concentration of α$'$ precipitates is known to increase continually with aging for as long as 4 years at 713 K [41]. This effectively means that the diffusion of Cr atoms remain active all along and so are the point defects that are known to assist solute atom movement in the lattice [42].

In view of this, we focus on the temporal evolution of α$'$ precipitates in conjunction with mapping the point defects at every stage of phase separation during thermal aging of binary Fe-20 at.% Cr alloy. A combination of techniques having atomic order resolution viz., atom probe tomography (APT) and positron annihilation spectroscopy (PAS), enables us to probe at the small spatial scales involved. APT is an established powerful microscopy and microanalysis technique that provides three-dimensional position and chemical identity of atoms [43]. Whereas, PAS is the only technique capable of detecting atomic order defects in concentration as low as a few ppm [44,45]. Individually, both these techniques are extensively used to investigate nano-scale precipitation and its correlation with age-hardening behaviour in various metallic alloys, often at the early stages. For example, APT is used for quantification of nano-scale α$'$ precipitates in terms of size, number density, volume fraction and chemical composition [32,37]. PAS, on the other hand, finds application in identifying point defects, solute clusters and second-phase nano-scale precipitates in alloys [4,46–50]. As for combining these two techniques, there are rather a limited number of studies and all of them are focused primarily on characterization of nano-scale precipitates and/or solute atom clusters under the effect of thermal aging and/or radiation [46,51–55]. However, the question of concomitant temporal evolution of point defects over the course of nano-scale precipitation remains largely unaddressed in the literature. Some of these questions can have far-reaching consequences. For example, (i) how do the nature and the characteristics of the associated point defects change as the solute clusters make transition to nano-scale precipitates, with increasing aging duration?; (ii) how do the associated point defects evolve as the nano-scale precipitates go through growth and coarsening stages, especially when a non-classical nature of nucleation-growth is operative? Understandably, one of the main hurdles faced by such investigation is the experimental difficulty involved in probing point defects and the nano-scale precipitates simultaneously at different stages of evolution.

This is where α$^/$ precipitates in Fe-Cr alloy system is quite special, and it offers an opportunity to follow the evolution of point defects in conjunction with the evolution of α$^/$-phase over a long span of aging duration. This can be explained in terms of the relative difference in the positron affinity between Cr and Fe. Due to the lower positron affinity of Cr with respect to Fe [56], a defect-free Cr-rich nano-phase (e.g., α$^/$) in Fe matrix is not an efficient trapping site of positron, and is indiscernible by PAS. These Cr-rich nano-phases become traceable by positron only when they are associated with point defects. Thus, the deciding factor is the association of point defects with α$^/$-precipitates (Cr-rich phase), which enables its detection by PAS. We, therefore, have been able to probe and fully characterize α$^/$ precipitates (Cr-rich phase) as well as the associated point defects at each stage of aging using APT and PAS, respectively. This investigation is performed on Fe-20 at.% Cr alloy that is known to undergo phase separation under thermal aging in non-classical nucleation-growth mode [32]. Thermal aging for this study is carried out at 773 K for durations spanning 1 to 1000 h and is intercepted after regular intervals for the combinatorial analysis using APT and PAS. This analysis identifies the dynamically evolving role of point defects in α$^/$ phase separation with progress of thermal aging, e.g., formation of vacancy-solute clusters in as-solutionized state, onset and proliferation of pre-precipitate solute clusters laden with vacancy clusters, emergence of nano-scale α$^/$ precipitates containing vacancy clusters, followed by appearance of virtually defect-free α$^/$ precipitates. To the best of our knowledge, no such systematic experimental study on the temporal evolution of point defects accompanying phase separation is available in the literature. The only other related work is a recent phase field simulation study that explores the co-evolution of point defects and Cr-rich nano-phase ($\alpha'$ precipitates) in binary Fe-Cr alloy under the effect of radiation [57]. Interestingly, many of the significant findings of the present study agree well with the predictions of the phase field simulation study. In addition, the present study provides new insight into the active role that point defects play in shaping the process of $\alpha'$ phase separation throughout the entire length of aging.

## 2. Experimental details

The buttons of binary Fe-20 at.% Cr alloy were prepared by vacuum arc melting high purity (99.99 %) elements in appropriate proportions. Homogeneity was ascertained by re-melting the alloy multiple times. The arc-melted alloy was hot-rolled to reduce thickness. The rolled specimen was sealed in a quartz ampoule filled with He, solutionized at 1273 K for 24 h, and quenched in ice water. Specimens of dimension 10 mm x 10 mm x 1 mm were cut from the rolled strip by electro-discharge machining and subjected to thermal aging at 773 K for the time duration of 1-1000 h. This heat treatment was followed by ice quenching. Detailed characterization of these specimens has been carried out using Scanning Electron Microscopy (SEM), X-Ray Diffraction (XRD), Wavelength Dispersive X-ray Fluorescence Spectrometry (WDS-XRF). The Vickers micro-hardness test was conducted on polished surface under a load and dwell time of 100 gm and 15 s, respectively. The final value of micro-hardness for each specimen was determined by averaging 10 measurements at 10 different locations of the specimen surface.

APT specimens were prepared by cutting matchstick-like pins of square cross section of dimension 0.3 mm x 0.3 mm and 8 mm length from the bulk specimen by electro-discharge machining. A standard 2-stage electro-polishing technique was followed to sharpen those pins into needles of suitable end radius (~ 50 nm). The initial coarse polishing was carried out with 10 vol. % perchloric acid in glacial acetic acid followed by fine polishing with 2 vol. % perchloric acid in 2-butoxyethanol. APT experiments were conducted using a Cameca[TM] FlexTAP instrument in laser-pulsing mode. The femto-second UV laser pulse energy was set at 30 nJ with pulse repetition rate of 50 kHz. Diaphragm was set at 15$^o$ for better mass resolution. Vacuum was maintained at 7 x 10$^{-11}$ Torr and the specimen temperature was 30 ± 0.5 K. Data analysis was carried out with IVAS 3.8.0 software (Cameca[TM]).

Positron annihilation lifetime (PAL) measurements were carried out using fast-fast coincidence spectrometer coupled with two BaF$_2$ scintillation detectors. The time resolution of the spectrometer was 216 ps. The $^{22}$Na positron source with an activity of ~ 20 μCi and sealed in ~ 8 μm Kapton® foil was used for PAL measurements. At least 2 x 10$^6$ coincidence events were accumulated for each measurement. PAL measurements were performed at room temperature. The positron lifetime spectra were analyzed using PALSfit [58]. The source contribution arising from positron annihilation in positron source and Kapton® foil after taking into account backscattering contribution consisted of two components viz., 0.398 and 1.91 ns and relative intensities 15.56 % and 3.11%, respectively. All the spectra could be fitted with two lifetime components to obtain a minimum $\chi^2$ (≤ 1.02).

The coincidence Doppler broadening (CDB) measurements were carried out using two high purity Ge (HPGe) detectors in coincidence and having an energy resolution of ~ 1.2 keV at 514 keV. The detectors were positioned at 180$^0$ relative to each other at a distance of ~ 8 cm apart with specimen placed at the middle position. At least 10$^7$ counts were recorded for each two-dimensional spectrum, which was then converted to a one-dimensional Doppler profile. These profiles contained information about the momentum distribution of the annihilating electrons and represented as ratio with respect to momentum distribution of a reference. The CDB ratio curves were obtained by normalizing the momentum distribution of each specimen to that of well-annealed Fe. The shape of the ratio curve in the high momentum region (typically $> 7 \times 10^{-3}\ m_oc$) exhibits the characteristic feature of the element due to positron annihilation with the inner core shell electrons.

## 3. Results

*3.1. Microstructural characterization*

The solutionized specimen possesses single phase ferritic structure (bcc) as confirmed by SEM and XRD analyses. No reflection corresponding to either σ phase (tetragonal) or the quenched-in austenite (fcc) phase is detected. Aging induced formation of α$^/$ phase is also not detected by XRD analysis since α$^/$ phase has the same bcc crystal structure and very similar lattice parameter like the matrix [59]. The chemical composition of the solutionized bulk specimen as determined by WDS-XRF is presented in Table 1. The composition is found to be very close to the nominal composition. The errors presented in the composition analysis correspond to one standard deviation.

*3.2. Micro-hardness measurements*

Aging induced change in the hardness of the alloy has been monitored through Vickers micro-hardness test. The variation of micro-hardness with aging time for Fe-20 at.% Cr is depicted in Fig. 1. The micro-hardness increases rapidly up to 120 h aging and the rate of increase becomes rather gradual afterwards, all the way up to 1000 h. Particularly striking is the sharp increase in micro-hardness at very early stage of aging (up to 25 h) before formation of any perceptible α$^/$ precipitates. Pre-precipitate clustering of Cr appears to be the reason and the next section analyzes this phenomenon in details. The magnified portion of the plot up to 50 h is shown as an inset of Fig. 1.

*3.3. APT measurements*

In order to find out possible clustering of Cr atoms during the initial stages of aging, a detailed investigation using APT has been carried out for the solutionized specimen and 1-50 h aged specimens. Fig. 2(a-d) show the Cr atom maps for the solutionized, 1 h, 25 h and 50 h aged specimens. The Cr-rich α′ precipitates are visible only in case of the 50 h aged specimen. Frequency distribution analysis (FDA) of the datasets is carried out with analysis volume of 20 nm x 20 nm x 50 nm and using a block size of 50 atoms. The result of the FDA is presented as the inset at the upper right corner of the respective atom map in Fig. 2(a-d). The experimental distribution matches well with the random binomial distribution for the solutionized specimen, but the aged specimens show deviations. The extent of deviation increases with aging time and the same has been quantified by the statistical analysis using $\chi^2$ and Pearson coefficient ($\mu$) [43]. The detailed results are presented in Table 2. Very small $\chi^2$ and $\mu$ values of 2.56 and 0.024, respectively indicate that the solutionized specimen is indeed random. On the other hand, steady increase of $\chi^2$ and $\mu$ values for the aged specimens suggests possible progressive formation of nano-structures. The very large $\chi^2$ and $\mu$ values ca., 3166.27 and 0.8554, respectively for the 50 h aged specimen suggest complete phase separation, discernible even from Cr atom maps. In addition, radial distribution function (RDF) analysis has been implemented to confirm the presence of Cr-rich clusters in the microstructure. In the RDF analysis, the average radial concentration of Cr is determined from each detected Cr atoms and normalized to the bulk Cr concentration. Mathematically, RDF can be written as [60–62]:

$$RDF = \frac{C_E(R)}{C_0} = \frac{N_E(R)/N(R)}{C_0} \qquad (1)$$

where $C_E(R)$ is the atomic composition of element $E$ at a distance $R$, $C_0$ is the average composition of element $E$ in the analyzed volume, $N_E(R)$ is the total number of atoms of all elements at the distance $R$. A step size of 0.1 nm was employed for the RDF analysis. The result is presented in Fig. 3 for 1 h, 25 h and 50 h aged specimens along with the solutionized specimen. The solutionized specimen shows a bulk normalized Cr concentration value close to 1, which increases as the aging progresses. The value above unity originates from positive interaction of Cr atoms and signifies aggregation of Cr. For instance, 50 h aged specimen shows the RDF value of 1.24 at $R = 0.1$ nm and Cr-rich $\alpha'$ precipitates are visible in the atom map, as shown in Fig. 2(d). On the other hand, 1 h and 25 h aged specimens show the RDF values of 1.04 and 1.11, respectively, suggesting Cr clustering, but these clusters are not visible in the atom maps (Fig. 2(b-c)). These results motivate us to carry out cluster search analysis for 1 h and 25 h aged specimens.

Quantitative analysis of clusters has been carried out using the maximum separation based cluster identification algorithm [63], as available in IVAS. The analysis is performed with an analysis volume of 20 nm x 20 nm x 50 nm of uniform density. First, cluster count distribution has been generated and the result is displayed along with the randomized data in Fig. 4(a-b) for 1 h and 25 h specimens, respectively and a $D_{max}$ value of 0.25 nm (dashed black line) is chosen. In the next step, $N_{min}$ is chosen from the cluster size distribution utilizing the $D_{max}$ value as obtained from the cluster count distribution. Fig. 5(a-b) shows the cluster size distribution along with the randomized data for 1 h and 25 h specimens, respectively. The dashed black line shows a reasonable estimate for $N_{min}$ of 20 atoms when compared to the randomized data. The matrix atoms in the cluster are accounted for by the envelope parameter $L$, and are removed from the cluster by the erosion parameter $E$ [63]. In this work, $L$ and $E$ values are taken as same as the $D_{max}$ value, i.e. 0.25. The analysis yields 115 clusters for the 25 h and 20 clusters for the 1 h aged specimens, Fig. 6(a-b), as demonstrated for one such specimen volume analysis. The spherical equivalent radius ($r$) of each cluster can be written as follows:

$$r = \sqrt[3]{\frac{3nV}{4\pi\eta}} \qquad (2)$$

where $V$ is the atomic volume ($V = \frac{a^3}{2}$), $a$ is the lattice parameter of Fe i.e. 0.2866 nm, $n$ is the number of atoms in the cluster and $\eta$ is the detector efficiency (0.62 for FlexTap) [37]. The cluster size (radius) distribution is presented in Fig. 7(a-b) as histograms. The average radius of the cluster is found to be 0.48 ± 0.03 nm for the 1 h and 0.52 ± 0.1 nm for the 25 h aged specimen. The number density of the clusters is determined by dividing the number of clusters by the analysis volume and is found to be $1 \times 10^{24} \pm 0.2 \times 10^{23}$/m$^3$ and $5.75 \times 10^{24} \pm 0.3 \times 10^{23}$/m$^3$ for the 1 h and the 25 h aged specimens, respectively. The volume fraction of the cluster is determined by the ratio between the number of Cr atoms present in the cluster and the overall solute atoms present in the analysis volume. The values of the volume fraction are found to be 0.225 ± 0.011 % for the 1 h and 1.72 ± 0.2 % for the 25 h aged specimens. Fig. 8(a-b) shows the Cr-concentration of the clusters as a function of their radius (as obtained from Eq. (2)) for 1 h and 25 h aged specimens. The average Cr concentration in the cluster for 1 h and 25 h aged specimens is estimated to be 58.22 ± 6.11 at.% and 59.16 ± 7.14 at.%, respectively.

As mentioned before, α$^/$ precipitates are detected first after 50 h aging. For all the aging durations from 50 h to 1000 h, detailed quantitative analysis of α$^/$ precipitates is carried out by APT using iso-concentration based method [64]. The Cr-threshold value for the analysis is arrived at based on the benchmark set by the size information obtained from TEM (transmission electron microscope), details of which can be found in our earlier work [65]. Additionally, SANS (small angle neutron scattering) analysis has validated the present APT-derived results [64,65]. Fig. 9 shows the temporal evolution of number density, size, volume fraction and composition of both the Cr clusters and α$^/$ precipitates, as obtained from APT analysis, spanning across the entire aging durations.

Right from 50 h onwards, precipitate size vs. time relationship follows $\sim t^{1/3}$ (inset of Fig. 9(a)), in line with Lifshitz-Slyozov-Wagner (LSW) theory of coarsening [66,67]. The number density (Fig. 9(b)) of the clusters increases up to 25 h, but that of the α$^/$ precipitates decreases with aging time from 50 h onwards. The slope (shown as a dashed line) of the number density vs time plot attains value ca. -1, as expected from LSW theory, only at higher aging times (750 h onwards). However, the volume fraction (Fig. 9(c)) increases continuously with aging time and does not saturate even after 1000 h of aging. Such time evolution of size, number density and volume fraction clearly indicates a non-classical process and is termed as transient coarsening, which arises due to the superposition of nucleation-growth with coarsening regime [68–70]. The non-classical nucleation is also evident from the Cr concentration profiles (Fig. 9(d)). It shows that the Cr concentration of α$^/$ increases continuously with aging time, while the matrix Cr concentration decreases at the same time, and the process does not saturate even after 1000 h of aging. The Cr concentration of α$^/$ observed in the current work for 1000 h aged specimen is 84.8 at.%, while the equilibrium Cr concentration of ~ 88 at.% is reportedly reached after 4 years at 773 K [41]. In addition, as noted before, the composition evolution of the Cr-rich clusters is found to be consistent with the temporal evolution of α$^/$ composition.

The APT analysis reveals early onset of phase separation well before the physical appearance of α$^/$ precipitates as apparent in the APT atom map. The Cr-rich clusters are precursors to α$^/$ precipitates in Fe-Cr alloys. The following section uses PAS to explore the evolving role of point defects as the pre-precipitate clusters emerge from the solutionized state and then make transition to α$^/$ precipitates and go through transient coarsening.

*3.4. Positron annihilation analysis*

The measured positron lifetime spectra in the studied Fe-Cr alloy (solutionized and aged specimens) could be decomposed into two components viz. $\tau_1 \approx 0.106$-$0.118$ ns and $\tau_2 \approx 0.25$-$0.28$ ns (with average standard error $\sim \pm 0.0014$ and $0.0053$ ns, respectively). These lifetimes represent different positron annihilation sites in the alloy. The shorter lifetime component ($\tau_1$) is remarkably lower than the lifetime in the smallest vacancy-like defect (i.e., monovacancy) in Fe and Cr (Table 3). Therefore, this component is attributed to free positron annihilation from delocalized state in the alloy matrix. The longer lifetime component ($\tau_2$) is attributed to the trapped state of positron. The theoretically calculated positron lifetime in vacancy-like defects in Fe ranges from $\sim 0.21$-$0.32$ ns for $V_2$-$V_{10}$ vacancy clusters [48,71–74]. A comparison with theoretical values of positron lifetime indicates that positron annihilates in vacancy-like defects (vacancy clusters). The measured positron lifetimes in solutionized and aged specimens are shown in Fig. 10. There is no appreciable difference between the measured $\tau_1$ in solutionized and aged specimens while, $\tau_2$ shows a significant change over the aging period. Above 50 h aging, $\tau_2$ increases monotonically up to 500 h from its value $\sim 0.250$ ns at 50 h to $\sim 0.297$ ns at 500 h. There is a marginal decrease in $\tau_2$ at 1000 h aging. However, the change in $\tau_2$ below 25 h aging is quite the opposite in nature (no significant change in $\tau_2$ between 25 and 50 h aging). The $\tau_2$ decreases with aging from 1 h to 25 h. There is only a marginal change in $\tau_2$ during the first 1 h of aging as compared to the solutionized state. Beyond that, $\tau_2$ decreases to $\sim 0.253$ ns at 25 h aging. Here we may note that chi-square reduced to number of degrees of freedom amounts to 1.05 and 1.10 at 25 h aging with $\tau_2$ as free parameter and fixed to $\sim 0.275$ ns (as for solutionized specimen), respectively. The substantial increase in chi-square when $\tau_2$ is fixed at 0.275 ns indicates that the decrease in $\tau_2$ for 25 h aged specimen is statistically significant.

The relative intensities of the longer positron lifetime component i.e., $I_2$ corresponding to the trapped state of positron in solutionized and aged specimens are shown in Fig. 11 (upper panel). The figure also presents in the lower panel the fraction of Cr atoms ($I_{Cr}$) surrounding the positron annihilation site (i.e., defects) as obtained from CDB, which will be discussed later. The measured $I_2$ varies in the range ~ 20-45 % over the studied aging period. The relative intensities of the two lifetime components indicate that a large fraction of positrons annihilate from free state in the bulk alloy matrix. Similar to $\tau_2$, the changes in $I_2$ are seen to be non-monotonic over the whole range of aging period. The $I_2$ increases sharply during the first 25 h of aging, and subsequently decreases gradually up to 1000 h of aging. It is interesting to note that $I_2$ increases appreciably during the first 1 h of aging as compared to its value in the solutionized state. The $I_2$ increases from ~ 25 % to ~ 35 % at 1 h aging indicating increase in the number of positron trapping sites or defects as soon as aging begins and continues up to 25 h. Beyond 25 h of aging, number of defects are seen to decrease as reflected from the decrease of $I_2$ for aging periods > 25 h. It is interesting to note that there is a substantial decrease in $I_2$ for 1000 h aged specimen ($I_2$ declines by ~ 50 % from its highest value at 25 h) and becomes even less than that in the solutionized specimen.

The measured positron lifetime spectrum in the studied alloy where α-α$'$ phase separation occurs represents positron annihilation in its trapped state in addition to annihilation in the bulk matrix. The different positron trapping sites can be: (i) vacancies associated with solute atom (vacancy-solute complex); (ii) vacancies trapped in solute clusters/precipitates; (iii) vacancies present in the alloy matrix outside clusters/precipitates; (iv) vacancies trapped at the interface of precipitate and the matrix.

In order to identify the nature of positron trapping site, its chemical surrounding has been examined using CDB. Fig. 12 shows the experimentally measured CDB ratio curves (normalized with respect to pure Fe) for specimens aged from 1 to 1000 h along with the ratio curves of pure Cr and solutionized specimen. The CDB ratio curve of pure Cr shows a broad peak at around 15 x$10^{-3}$ $m_o c$ unit, which is the characteristic feature of Cr in the ratio curve.

The ratio curve of solutionized specimen shows only a marginal deviation from unity with only a subdued feature of Cr. This feature of the curve indicates that positrons predominantly annihilate with Fe atoms in the solutionized specimen. With aging, the ratio curves show a systematic increase in the amplitude of the characteristic peak of Cr up to 25 h where the peak amplitude reaches a maximum. Above 25 h aging, the peak amplitude decreases systematically up to 1000 h. For 1000 h aged specimen, the ratio curve is almost identical to that of pure Fe suggesting annihilation predominantly with Fe atoms. It is seen that the amplitude of Cr peak in the ratio curves doesn't overlap with pure Cr, which suggests that all the atoms around the annihilation site are not of Cr and the change in the amplitude of Cr peak shows that fraction of Cr atoms around annihilation sites varies with aging. We have estimated the fraction of Cr atoms ($I_{Cr}$) around the annihilation site from the amplitude of the broad peak in the ratio curve, taking account of the annihilation from inner core electrons [75]. The position of each point on the ratio curve spans the range between pure Fe ($I_{Cr} = 0$) and pure Cr ($I_{Cr} = 100\%$). The variation of $I_{Cr}$ with aging is shown in the lower panel of Fig. 11. Interestingly, the variation of $I_{Cr}$ follows $I_2$ over the whole aging period showing increasing trend up to 25 h and a monotonic decrease with subsequent aging up to 1000 h. The change in $I_{Cr}$ from solutionized state to 25 h aging is rather sharp. The $I_{Cr}$ increases from ~ 25% to 58% from the solutionized state to 1 h of aging and reaches a maximum (~ 70%) at 25 h of aging. At the highest aging time i.e. 1000 h, $I_{Cr}$ reduces to ~ 15%, which is even less than the solutionized state.

The isolated Cr atoms and/or, defect-free Cr-rich phase embedded in Fe matrix are not efficient trapping site for positrons owing to the lower positron affinity of Cr in comparison to Fe [56]. The positron affinity of Cr and Fe are -2.62 and -3.84 eV, respectively. Hence, in the studied alloy, only the Cr-rich clusters associated with vacancy-like defects would contribute towards positron annihilation. This suggests that the measured $I_{Cr}$ is directly related/dependent on $I_2$ (a representative of concentration of point defects) and thereby, provides direct information about the chemical surrounding of defects present in the alloy. It is interesting to note that, $I_2$ is lower than $I_{Cr}$ for all the aged specimens except 1000 h aged specimen. Also, it is seen that $I_{Cr}$ increases much more than the increase in $I_2$ with aging. For instance, at 1 h of aging, $I_2$ increases to only ~ 36 % from its value in solutionized state while $I_{Cr}$ increases to ~ 58 %. The $I_{Cr}$ ~ 58% signifies that nearly 58% of the neighbouring atoms of the vacancy defect are of Cr. This fraction increases to about ~ 70 % for 25 h aged specimen whereas $I_2$ has increased to only 43 % at this aging. Although increase in $I_2$ would lead to increase in $I_{Cr}$, the observed magnitude of increase in $I_{Cr}$ as compared to $I_2$, suggests that with aging, Cr atoms aggregate towards vacancy sites. In the solutionized state, $I_{Cr}$ and $I_2$ are nearly the same while at the lowest aging studied i.e. 1 h, $I_{Cr}$ becomes nearly double in contrast to $I_2$. We, therefore, state that a large fraction of Cr atoms around the trapping site at 1 h aging signifies clustering of Cr atoms. Furthermore, the results indicate that number of Cr atoms in the cluster increases with aging up to 25 h. Eventually, the $I_{Cr}$ decreases monotonically following the trend of $I_2$ and for the highest aging time (i.e. 1000 h), it decreases to nearly 15%. Only at the highest aging time (1000 h), $I_{Cr}$ is seen to be less than $I_2$. It is interesting to note that PAS result corroborates with APT revelation of formation of Cr clusters at 1 h of aging.

## 4. Discussion

Phase separation in Fe-20 at.% Cr alloy follows a non-classical nucleation-growth mode with continually increasing Cr concentration of pre-precipitate clusters and α′ precipitates throughout the aging duration of 1000 h. In this context, special mention should be made of the article by Nie et al. that critically examines the nature and characteristics of the pre-precipitate clusters vis-a-vis the precipitates for a wide variety of systems [14]. A pre-precipitate cluster can be envisaged as a Cr-rich atomic aggregate without an independent crystal structure and therefore, coherent with the matrix [14,76]. It is the formation of $\alpha'$ crystal structure that marks the inception of $\alpha'$ precipitates and differentiates it from a pre-precipitate cluster [14]. But, experimentally isolating a pre-precipitate cluster from a nano-scale precipitate is extremely challenging, particularly so in case of Fe-Cr system. This is because both $\alpha'$ precipitate and $\alpha$ matrix have the same bcc crystal structure and the difference in their lattice parameter values is imperceptibly small [59]. APT appears to be the more realistic option in absence of any viable diffraction-based alternative. However, it should be mentioned here that isolating pre-precipitate clusters from the nano-scale $\alpha'$ precipitates by APT is done essentially on the basis of statistical correlation analysis and the size of the nano-scale feature and, therefore, can be subjective and contentious. The current study, aware of the uncertainty involved, makes a distinction between the $\alpha'$ precipitates and their pre-cursors on the basis of the APT analysis. However, as revealed by the PAS analysis, significant difference between (i) the natures of the point defects associated with them; and (ii) point defects evolution over the aging period, lends credence to this method of distinction. Based on the APT analysis, the first sign of phase separation in terms of formation of Cr clusters is noticed as early as after 1 h of aging at 773 K. These pre-precipitate clusters grow in numbers and size up to 25 h. The size of the Cr clusters grows from $0.48 \pm 0.03$ nm after 1 h to $0.52 \pm 0.1$ nm after 25 h, and the number density reaches the peak value of $5.75 \times 10^{24}/m^3$ after 25 h aging. This explains the rapid rise in the micro-hardness up to 25 h of aging. The increase in hardening due to the formation of solute clusters is reported in many other alloys and is referred to as cluster hardening [12,77–84].

50 h of aging onwards, experimental evidence of formation of $\alpha'$ precipitates is clearly demonstrated by the APT analysis. The average radius of $\alpha'$ precipitates is found to be $0.94 \pm 0.34$ nm at 50 h, which increases progressively to $2.35 \pm 0.43$ nm at 1000 h. The average Cr concentration of $\alpha'$ precipitates increases from $60.1 \pm 0.4$ at.% at 50 h to $84.8 \pm 0.5$ at.% at 1000 h aging. In fact, the evolution of size, number density and volume fraction clearly indicates a transient coarsening, which arises due to the superposition of nucleation with coarsening region [68–70]. As mentioned earlier, the non-classical nucleation entails a critical role of diffusion of Cr atoms and in turn that of the point defects like vacancies in this phase separation process.

*4.2.1. Point defects in the solutionized specimen*

Previous studies have reported the occurrence of solute-rich composition fluctuations and/or, solute clusters during or immediately after quenching from the solutionizing temperature that affect the aging behaviour of the alloy [85,86]. Our results (PAS and APT), however, do not indicate the presence of clusters and/or, solute-rich composition fluctuation in the solutionized specimen. However, PAS results suggest presence of vacancy-like defects in the solutionized specimen as indicated by positron annihilation parameters viz., $\tau_2$ and $I_2$.

In general, vacancies are inherent to materials, especially metals and alloys. The concentration of vacancies at a given temperature can be estimated using thermodynamic analysis as well as from the positron data. The concentration of equilibrium thermal vacancies ($C_v^*$) in the solutionized specimen at $T$ = 1273 K can be approximated by:

$$C_v^* \approx exp\left(-\frac{E_v^f}{kT}\right) \qquad (3)$$

where $E_v^f$ denotes the vacancy formation energy and $k$ is Boltzmann constant. The vacancy formation energy in Fe-Cr alloy is reported to be dependent on the Cr content as well as the configuration of vacancy [87]. Based on reference [87], we assume the average $E_v^f$ ~ 1.83 eV and estimated $C_v^*$ is ~ 1.1 x10$^{-8}$.

The concentration of vacancies estimated from positron lifetimes and intensities using the two-state standard trapping model [77,78] is given by:

$$C_v^{equilibrium} = \frac{1}{\mu_v}\frac{I_2}{I_1}\left(\frac{1}{\tau_b}-\frac{1}{\tau_2}\right) \quad (4)$$

where $\tau_b$ is the bulk positron lifetime of the alloy, $\mu_v = i \times \mu_1$ [88]; $\mu_v$ and $\mu_1$ are specific positron trapping coefficients for vacancy cluster and monovacancy, respectively, $i$ is the number of vacancies in the vacancy cluster. Considering $\mu_1 \sim 1.1 \times 10^{15}$ s$^{-1}$ for monovacancy in Fe [71], the calculated $C_v^{equilibrium}$ is 2.38 x 10$^{-7}$.

The experimentally obtained vacancy concentration (from PAS, i.e., $C_v^{equilibrium}$) is seen to be higher than the equilibrium concentration of thermal vacancies ($C_v^*$) at solutionizing temperature ($T$ = 1273 K) indicating the presence of non-equilibrium thermal vacancies or excess vacancies in the solutionized specimen. The concentration of these excess vacancies depends on a number of factors like solutionizing temperature, quenching rate and the presence of solute atoms [89–91]. The stability of these vacancies is generally ascribed to the formation of vacancy-solute (V-S) complexes. The studied Fe-Cr alloy has Mn (0.2 at.%) and Si (0.1 at.%) in addition to Cr (20.5 at.%) as seen from the APT based composition analysis. The formation of V-S complex (S: Si, Mn, Cr) depends on the binding energy of the V-S complex and the migration energy of the solute in Fe. *Ab initio* calculations have shown that Mn has high migration energy in Fe and weak binding energy with the vacancy [92]. The combination of high migration energy and weak binding energy suggests formation of V-Mn to be highly improbable. In contrast, binding energies of V-Si is strong enough to form V-S complex, but the dependence on magnetic ordering and migration energy in Fe matrix makes the situation quite complicated for this solute [93]. However, there is a finite probability of formation of V-Si in the studied alloy. For V-Cr complex, the binding energy changes with the size of V-Cr complex as well as its geometric configuration [93]. Lavrentiev et al. [94] have done a rigorous analysis of V-Cr complexes in dilute Fe-Cr alloy using *ab initio* calculation and showed that the binding energy of V-Cr complex increases drastically with the increase in the number of Cr atoms in the complex forming stable V-Cr complex. It is reported that for the first nearest neighbour configuration the binding energy is negative and it increases when Cr atoms are at higher order of near neighbour distances [94]. All these studies have shown the stability of V-Cr complex in Fe-Cr alloys. Considering a high Cr content in the studied alloy, the formation of V-Cr complexes is highly probable during quenching following the solutionizing treatment. Ignoring the small fraction of V-Si in contrast to V-Cr complex, we ascribe point defects present in solutionized specimen to V-Cr complex originated from non-equilibrium thermal vacancies at the solutionizing temperature.

*4.2.2. Point defects in the aged specimens: Cr clustering (≤ 25 h of aging)*

The presence of point defects in the aged specimens is revealed from trapped positron annihilation parameters viz. lifetime ($\tau_2$) and its corresponding intensity ($I_2$), and the corresponding chemical surrounding of the defects from $I_{Cr}$. The sharp increase in $I_2$ from the solutionized state to 25 h aging indicates increase in the number of defects. In addition, $I_{Cr}$ indicates that defects are predominantly decorated with Cr atoms. Up to 25 h of aging, the fraction of Cr atoms around the defects increases, which is indicative of Cr aggregation around the defects and formation of Cr clusters. This is consistent with our APT analysis that shows initiation of Cr clustering at the lowest aging time i.e., 1 h and increase in the number density as well as the size of Cr clusters up to 25 h. The combined APT and PAS results, therefore, suggest that Cr clustering is associated with increase in the number of defects. We may note that the increase in $I_2$ from the solutionized state to 1 h aging is quite appreciable, which suggests that these defects are not thermally generated. The equilibrium concentration of thermal vacancies is much lower at the aging temperature (773 K) than the solutionizing temperature (1273 K). Assuming the presence of excess vacancies (present in the solutionized specimen as discussed in the previous section) and thermally generated vacancies at the aging temperature, the observed $I_2$ at 1 h aging still cannot be accounted for. Therefore, it is strongly suggested that aging induced Cr clustering is responsible for generation of point defects in this case. These defects are vacancy clusters trapped in Cr clusters as inferred from the measured $I_{Cr}$. The increase in $I_2$ up to 25 h further confirms that defects are generated during Cr clustering. Interestingly, $I_{Cr}$ follows $I_2$ but increases at a faster rate than $I_2$, indicating that size of Cr clusters increases from 1 h to 25 h, which is consistent with the APT result.

As stated earlier, the defects at this stage are primarily clusters of vacancies trapped in the Cr clusters. However, based on the variation of the measured $\tau_2$ between 1 h and 25 h, it is difficult to conclusively determine any change in the size of these defects. Although, measured $\tau_2$ suggests that vacancy clusters are of about 4-6 vacancies, the association of vacancy with Cr atoms as well as arrangement of Cr atoms around the vacancy site can influence electron density distribution, which may be reflected in the measured positron lifetimes. The effect of Cr atoms near the vacancy site on structural relaxations of the atoms from the ideal lattice positions has been investigated using *ab initio* approach [94]. It has been stated that inward or outward relaxation of atoms from the vacancy depends on the nearest neighbour position of Fe and Cr atoms as well as their relative number. For instance, Cr atoms at first nearest neighbour position tend to relax inwardly towards the vacancy while at the second nearest neighbour position, an outward relaxation is predicted. It is likely that the observed variation of $\tau_2$ from 1 h to 25 h is a consequence of arrangement Cr atoms surrounding the vacancies, which influences the electron density distribution around the vacancy clusters.

*4.2.3. Point defects in the aged specimens: $\alpha'$ precipitates (> 50 h of aging)*

50 h onwards, $I_2$ and $I_{Cr}$ decrease with aging up to 1000 h. The decrease in $I_2$ indicates decrease in the number of defects in the alloy. It is interesting to note here that APT result shows that the number density of the Cr-rich α′ precipitates starts decreasing 50 h onwards (Fig. 9b). Our PAS result, therefore, suggests that point defects are generated only during initial clustering stage i.e. up to 25 h of aging. The transition of these clusters into the nano-scale α′ precipitates do not lead to further generation of defects. The aging period beyond 25 h (i.e., 50 h onwards) constitutes nucleation, growth and coarsening of α′ precipitates. The decrease in the number of defects is likely to be related to overall reduction in α′ precipitate number density and/or, recovery of defects trapped inside the precipitates during the transient coarsening process. We observe that decrease in $I_2$ from 25 h to 500 h is gradual but slow in contrast to $I_{Cr}$, which although follows $I_2$ but its rate of reduction is much faster, especially above 120 h aging. This indicates that the chemical surrounding of the point defects in regards to fraction of Cr atoms surrounding the defect or, in other word, the nature of point defects changes during this aging period. This implies that positron annihilates at sites other than the point defects trapped inside the Cr precipitates. We ascribe the slow decrease in $I_2$ and $I_{Cr}$ between 25 and 120 h primarily to the decrease in the number density as the Cr-rich clusters make transition to α′ precipitates. However, between 120 to 500 h, the sharp change in $I_{Cr}$ suggests that nature of defects, i.e., the chemical surrounding of the defects changes significantly, while there is no significant change in the number of defects. This can be explained on the basis of migration of defects from the precipitates to the matrix. The defects trapped in α′ precipitates have predominantly Cr atoms surroundings as compared to the defects present in the matrix where Fe atoms are also present around the defects. The sharp decrease in $I_{Cr}$, therefore, reveals recovery of defects from the precipitates and their migration to the matrix. It is to be noted that the number of defects during this period does not change appreciably as compared to change in the $I_{Cr}$. Hence, based on our APT and PAS results, we infer that during the transient coarsening of precipitates (i.e. 120 h to 500 h of aging), defects are recovered from the precipitates and moved to the matrix.

This suggests that α$^/$ precipitates grow in size through recovery of defects from the precipitates and their migration to the matrix. It, therefore, appears that during the aging period 120 h - 500 h, the matrix primarily consists of defect-free α$^/$ precipitates and vacancy clusters. However, the process is gradual and there is a finite probability of presence of Cr-rich α$^/$ precipitates that are not entirely defect-free. As shown before, Cr-enrichment of α$^/$ in Fe-20 at.% Cr is a continuous process and even after 1000 h aging the equilibrium concentration of Cr in precipitates is not achieved. Our PAS results show that at 1000 h, nearly all the precipitates become defect-free and the defects that have migrated to the matrix are also recovered, as indicated by $I_{Cr}$ and $I_2$. The effective concentration of these defects in the matrix is even less than the concentration of defects observed in the solutionized specimen. Our results suggest that the recovery of defects from the precipitates and further from the matrix is a continuous process.

We observe a systematic increase in $\tau_2$ during 50 to 500 h aging and a marginal decrease at 100 h of aging. Although the observed $\tau_2$ corresponds to vacancy clusters of 4-6 vacancies, the change in $\tau_2$ with aging cannot be unambiguously ascribed to change in vacancy cluster size. It is seen that $\tau_2$ increases during the cluster to precipitate transition (50 h onwards) and subsequent growth and coarsening stages of the precipitates. We suggest that structural rearrangement of atoms during precipitate formation and their growth and coarsening are responsible for the observed change in $\tau_2$. We may note here that $\tau_2$ increases significantly above 240 h, the period which we have ascribed to a region of recovery of defects from precipitates via migration to the matrix. A recent phase field simulation study has shown that during the growth process, point defects migrate from the initial cluster position i.e., core of the cluster to the interface of α-α$^/$ phases and accumulate into a defect concentration loop around α′ phase [57]. Even though the nano-scale α′ precipitates are known to remain coherent throughout the aging process [95,96], the migration of defects from precipitates interior to the matrix during the growth process appears to be a validation of the phase field simulation result. We admit that it is difficult to identify different type of defects and their exact locations during this stage and therefore, the formation of defect concentration loop cannot be experimentally verified. Our comprehensive APT and PAS analysis of aging kinetics in Fe-20 at.% Cr is an experimental account of co-evolution of point defects and Cr-rich nano-phase during a non-classical nucleation-growth process. The dynamics of vacancies during the process of non-classical nucleation-growth process of α′, starting from pre-precipitate clustering to the formation of α′ precipitates is explained pictorially in Fig. 13.

## 5. Conclusions

The current study presents the continually evolving dynamics of point defects that accompanies the non-classical nucleation-growth mode of α′ phase separation process during thermal aging of Fe-20 at.% Cr alloy at 773 K, starting from the solutionized condition and including the pre-precipitate clustering stage. While APT has monitored the progress of Cr-rich phase separation, PAS, because of the lower positron affinity of Cr with respect to Fe, has been successful in probing specifically the Cr-rich nano-phase associated with point defects at every stage of aging. Our results reveal for the first time the crucial role that the point defects play in mediating α′ phase separation in Fe-Cr system during the entire aging duration. The major findings are listed below:

1. The solutionized specimen shows the presence of vacancy-Cr complexes that subsequently act as nucleation sites for the pre-precipitate Cr-rich clusters.
2. Even though α′ precipitates are detected 50 h onwards, phase separation at 773 K in Fe-20 Cr at.% alloy is initiated through formation of pre-precipitate Cr-rich clusters as early as after 1 h of aging. APT analysis has successfully detected and quantified the pre-precipitate Cr-rich clusters that are responsible for the sharp increase in micro-hardness at the early stage (up to 25 h) through cluster hardening. The clustering of Cr atoms has also been corroborated by PAS.

3. Proliferation of the pre-precipitate Cr-rich clusters up to 25 h leads to an initial increase in the vacancy concentrations that are associated with the Cr-rich clusters. The vacancy concentrations keep declining once the Cr-rich clusters make transition to Cr-rich $\alpha'$ precipitates.

4. Our results reveal that with progress of phase separation, the point defects associated with $\alpha'$ precipitates migrate from the core of the precipitates towards the precipitate/matrix interface and eventually get recovered.

**Acknowledgements**

Authors would like to thank Mr. Santosh Yadav, Technician, MSD for his enormous help in every step of specimen preparation.

# Tables

## Table 1

| Sol. | Fe | Cr | Si | Mn | S | P |
|---|---|---|---|---|---|---|
| Bulk-WDS-XRF | 80.2±0.1 | 19.8±0.1 | ---- | ----- | <0.02 | <0.02 |
| Local-APT | 79.2±0.03 | 20.5±0.04 | 0.1±0.003 | 0.2±0.004 | ------ | ------ |

**Table 1.** Bulk chemical composition for solutionized Fe-20 at.% Cr alloy as obtained from WDS-XRF and APT respectively.

## Table 2

| Aging time (h) | Considered atom | Reduced $\chi^2$ | $n_d$ | $p$-value | $\mu$ |
|---|---|---|---|---|---|
| Sol. | Cr | 2.56 | 17 | <0.0001 | 0.0240 |
| 1 | Cr | 53.47 | 19 | <0.0001 | 0.2122 |
| 25 | Cr | 2731.90 | 20 | <0.0001 | 0.6475 |
| 50 | Cr | 3166.27 | 21 | <0.0001 | 0.8554 |

**Table 2.** Parameters obtained from $\chi^2$ statistical test of APT data as a function of aging time.

**Table 3**

| **Positron lifetimes (ns)** | | | | | | |
|---|---|---|---|---|---|---|
| | **Bulk** | **$V_1$** | **$V_2$** | **$V_4$** | **$V_6$** | **$V_{10}$** |
| **Fe** | 0.111 | 0.175 | 0.200 | 0.240 | 0.270 | 0.290 |
| **Cr** | 0.120 | 0.200 | 0.225 | 0.255 | 0.275 | 0.310 |

**Table 3.** Reported positron lifetimes in vacancy-clusters ($V_n$, *n* is the number of vacancies in the cluster) in Fe and Cr taken from references [71-74].

**Figures**

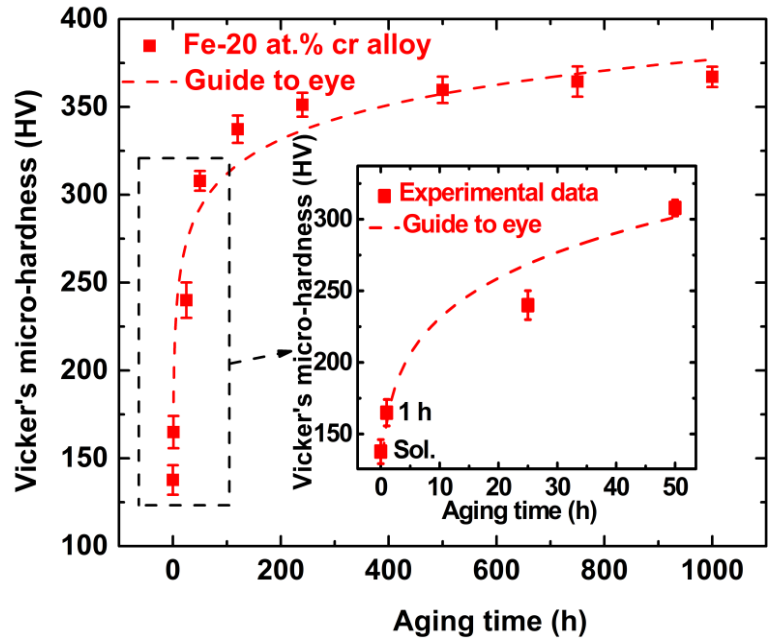

**Fig. 1.** Evolution of Vickers micro-hardness with aging time.

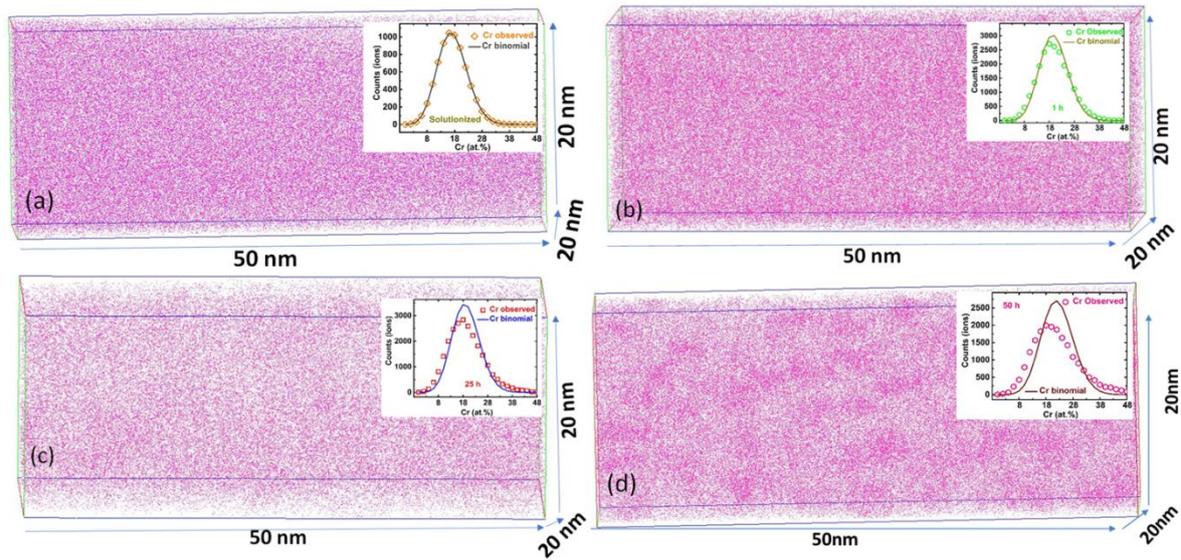

**Fig. 2.** Cr atom maps of binary Fe-20 at.% Cr alloy for: (a) solutionized condition, and aged at 773 K for time durations of: (b) 1 h, (c) 25 h and (d) 50 h respectively. Each purple dot represents one Cr atom. The inset shown in the upper right corners of each image corresponds comparative plots of observed Cr atom distribution with random binomial distribution.

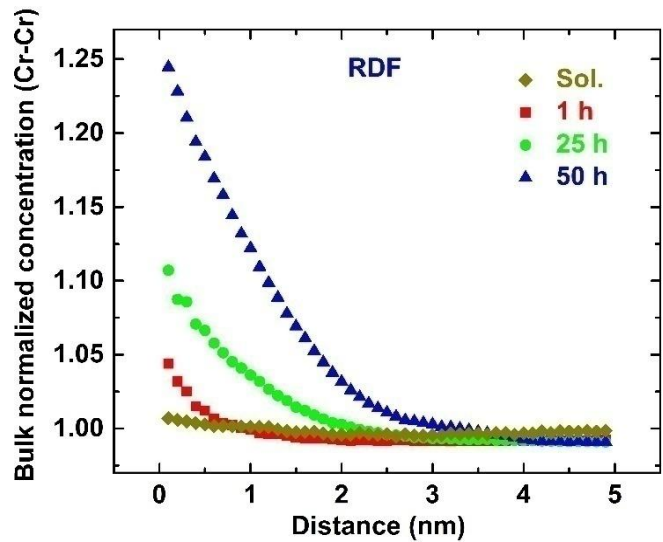

**Fig. 3.** Normalized RDF curves of Cr-Cr for solutionized and aged (1- 50 h) samples for Fe-20 at.% Cr alloy.

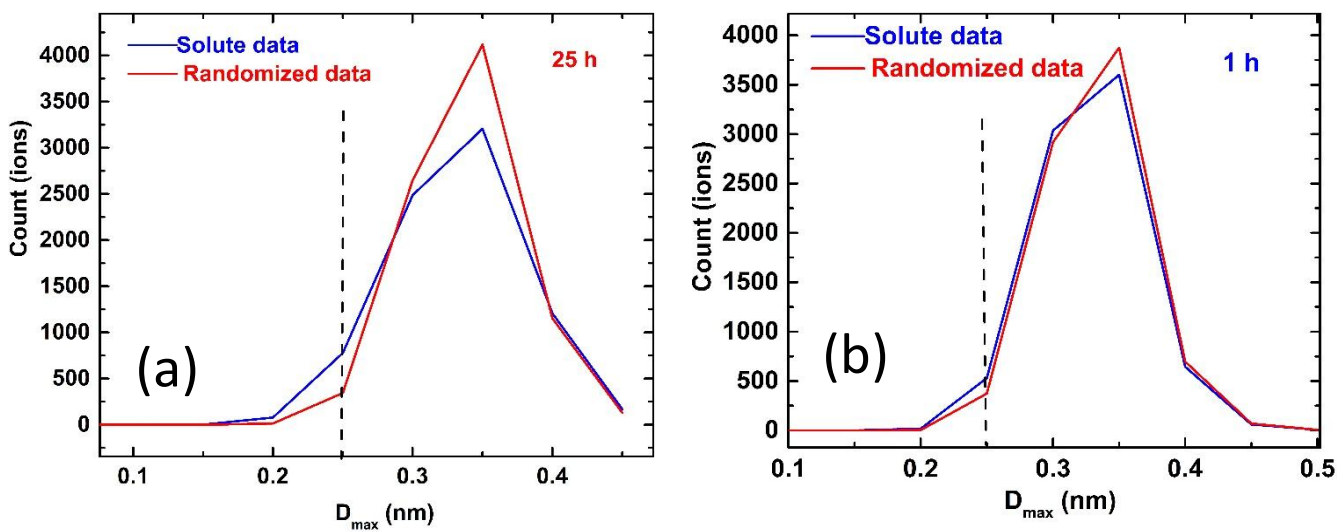

**Fig. 4.** Cluster count distribution for actual data (blue line) along with the data randomizing all ions (red line), as displayed for: (a) 1 h and (b) 25 h aged sample.

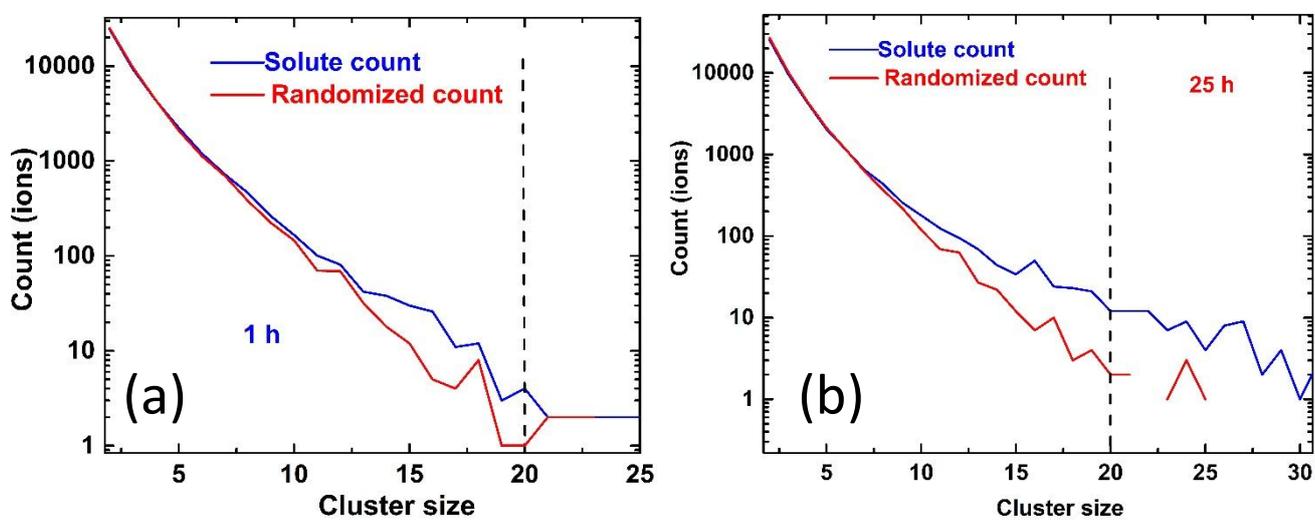

**Fig. 5.** Cluster size distribution using $D_{max}$ value of 0.25 for actual data (blue line) along with the data randomizing all ions (red line), as displayed for: (a) 1 h and (b) 25 h aged sample.

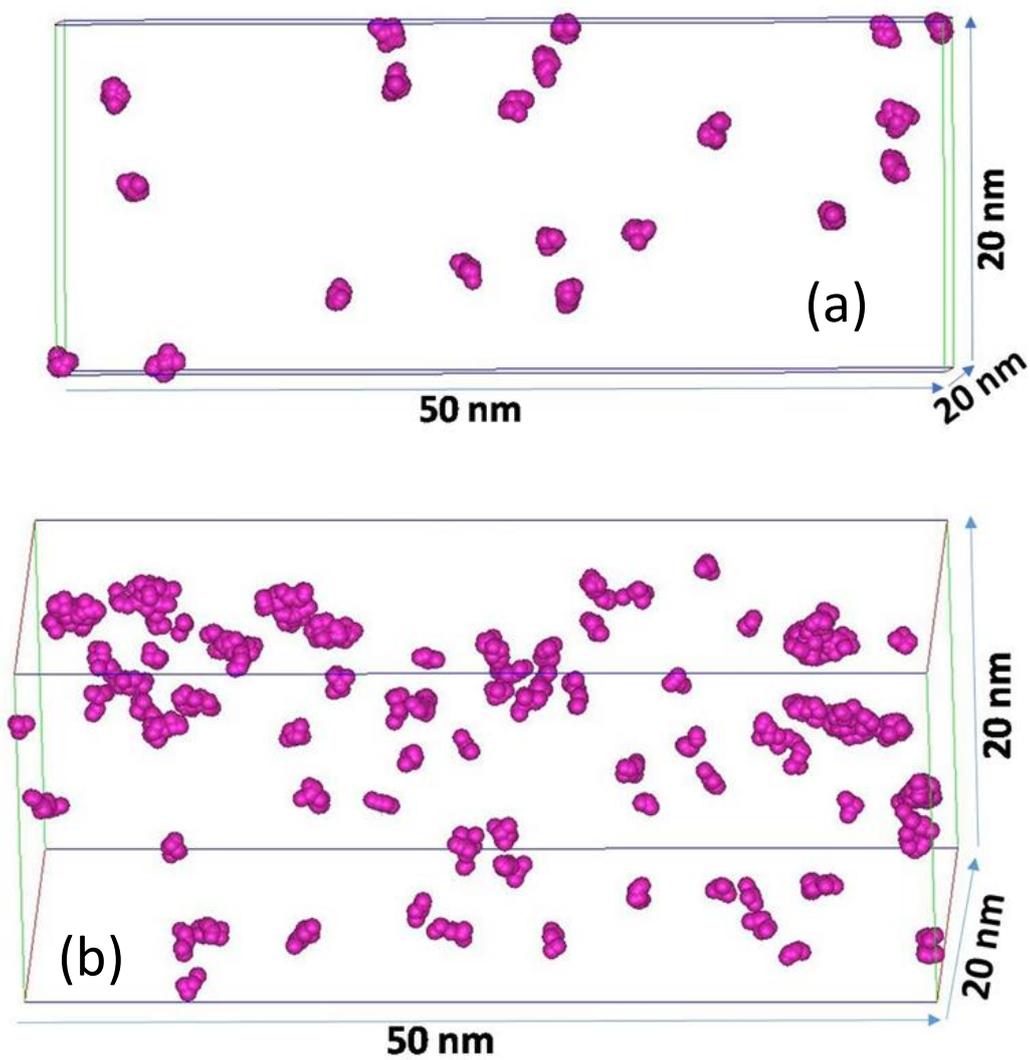

**Fig. 6.** Distribution of Cr clusters, obtained from cluster detection algorithm, in the analysis volume as displayed for: (a) 1 h and (b) 25 h aged sample.

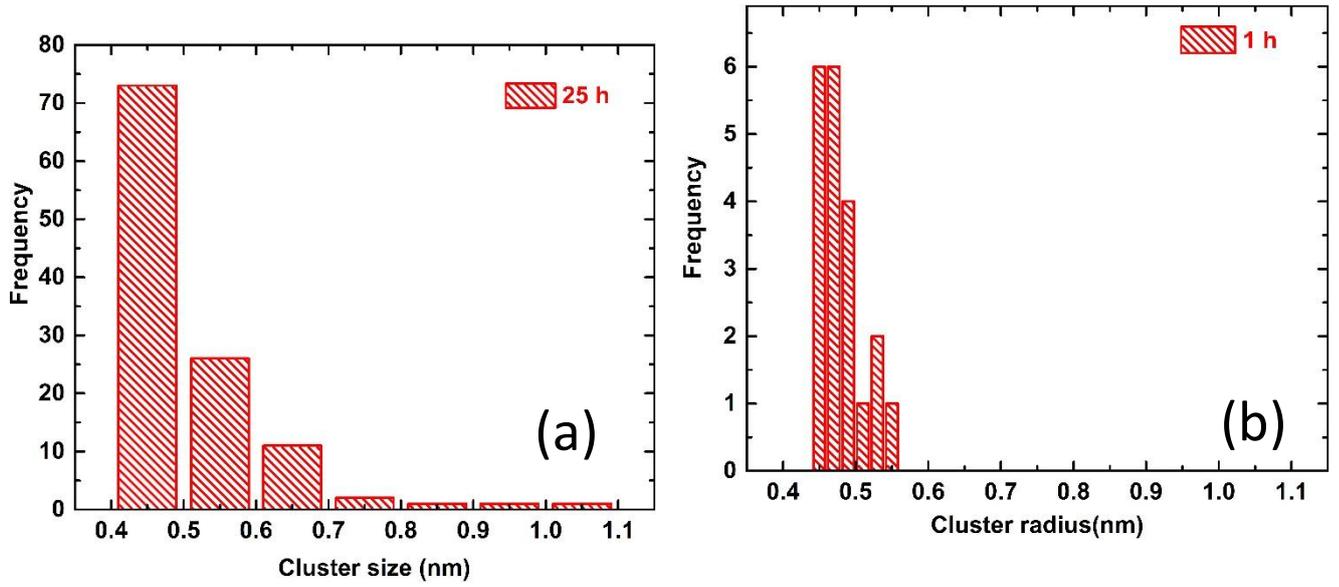

**Fig. 7.** Cluster size distribution shown as a histogram for: (a) 1 h and (b) 25 h aged sample.

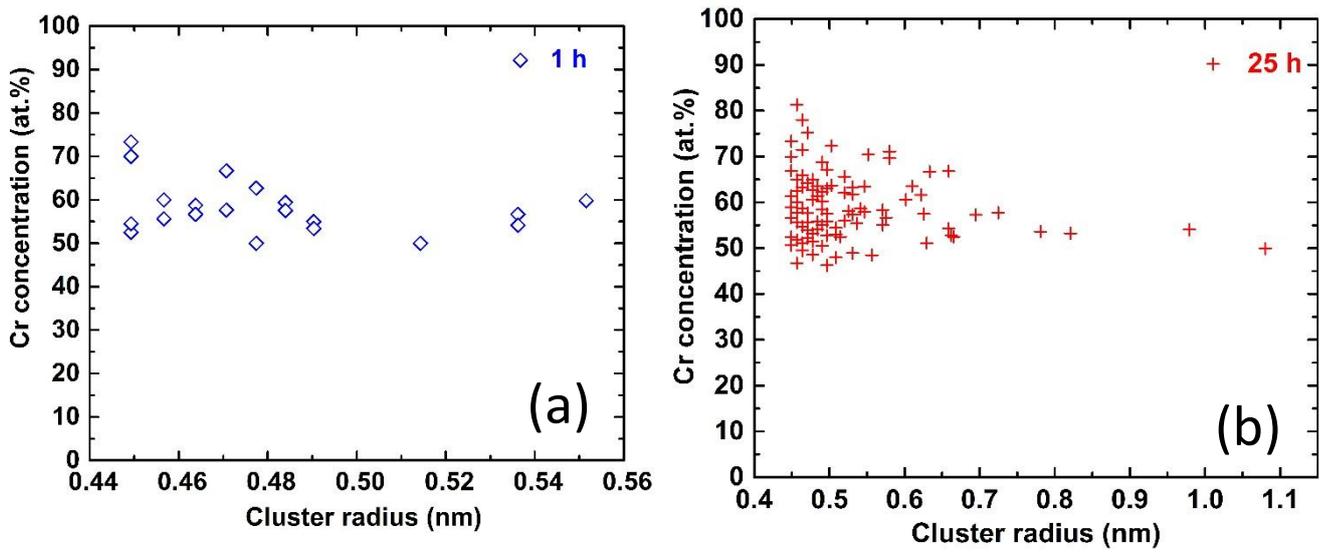

**Fig. 8.** Cr concentration of the clusters as a function of their radius, as shown for: (a) 1 h and (b) 25 h aged sample.

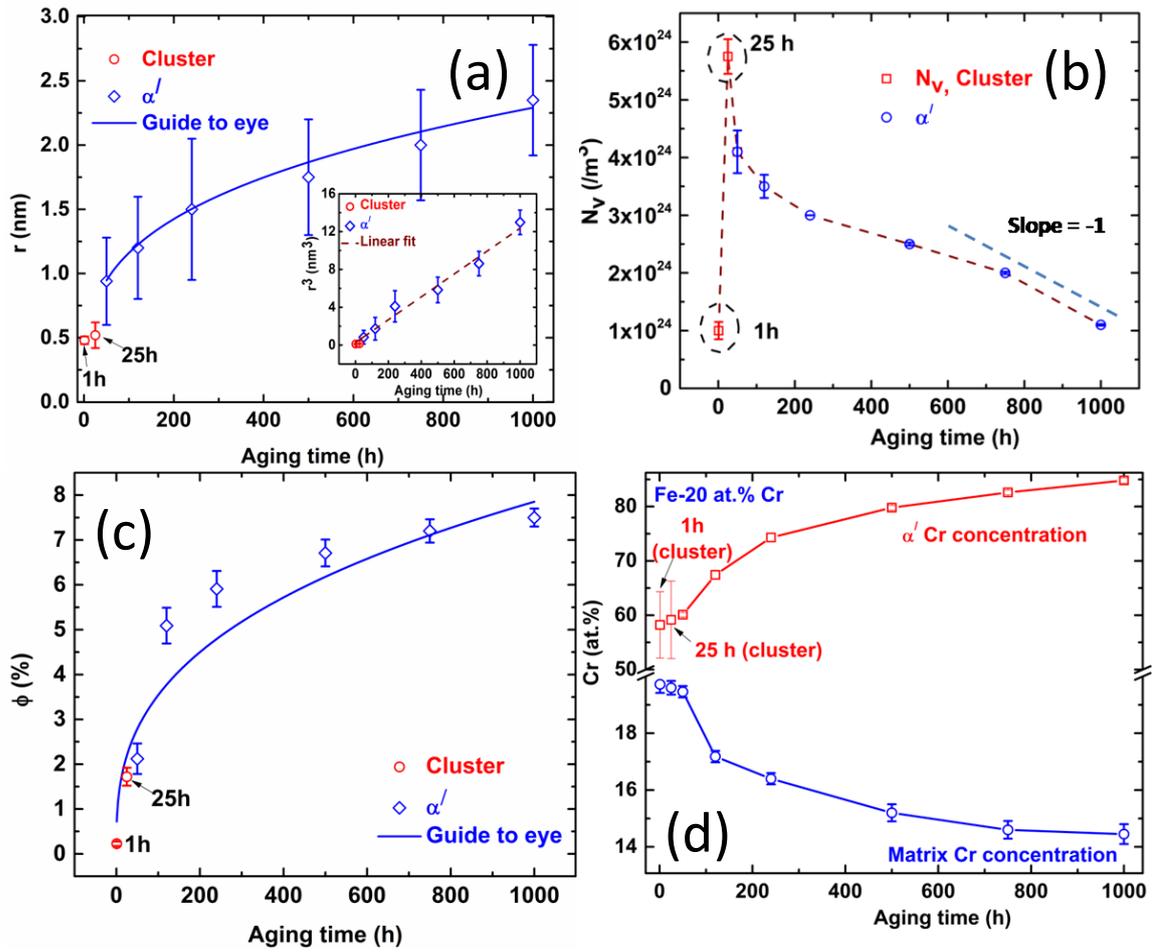

**Fig. 9.** Temporal evolution of: (a) average radius, (b) number density, (c) volume fraction and (d) Cr-concentration of Cr-rich nano-phase, while inset of Fig. (a) demonstrates $\sim t^{1/3}$ relationship of precipitate radius with time. The 50 h onwards data is taken from reference [64].

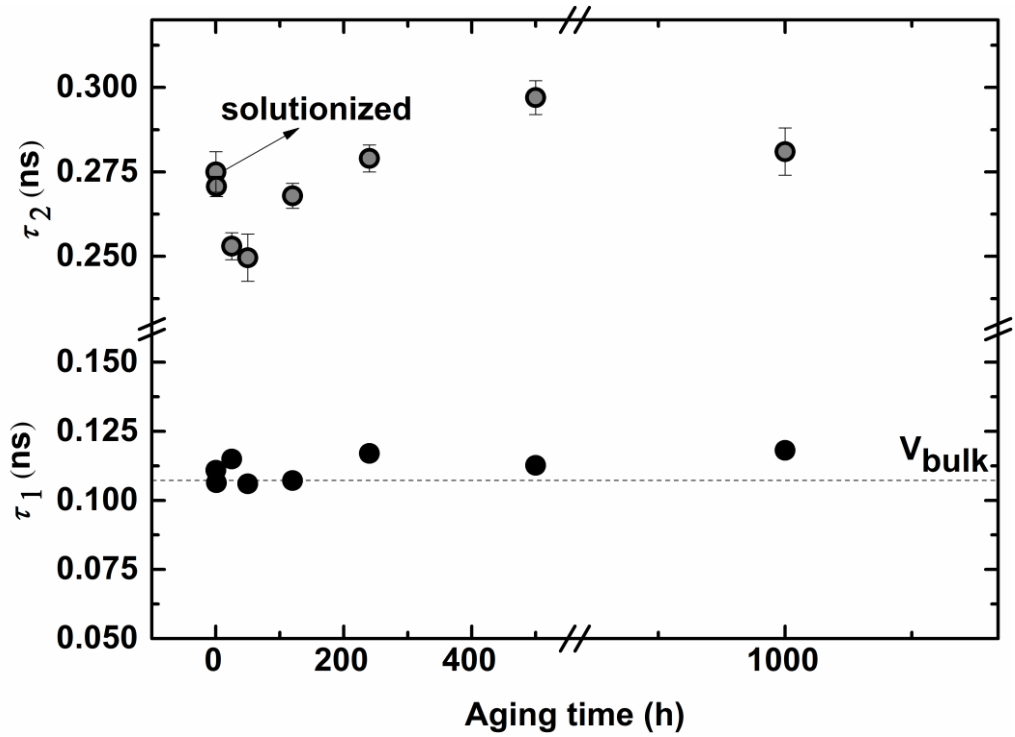

**Fig. 10.** The variation of positron lifetimes ($\tau_1$ and $\tau_2$) as a function of ageing time for Fe-20 at.% Cr alloy aged at 773 K. The lifetime components viz. $\tau_1$ and $\tau_2$ represent free and trapped positron annihilation, respectively in the alloy. The dashed lines represent theoretically calculated values of positron lifetimes in the bulk and monovacancy in pure Fe for reference [Table 3].

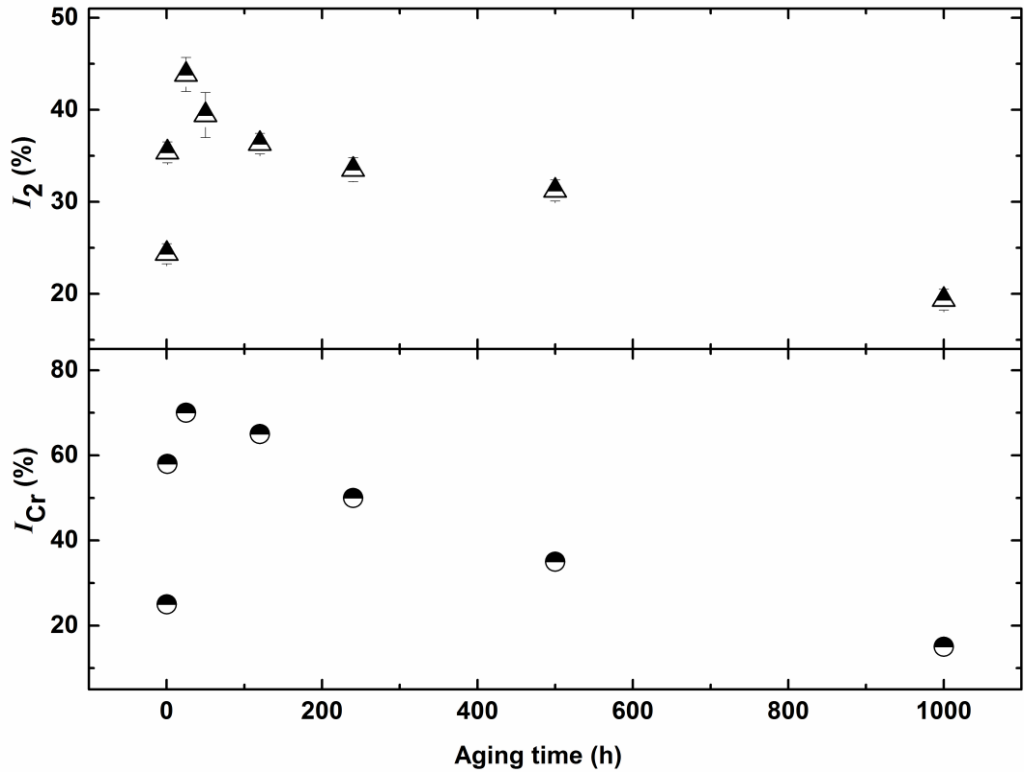

**Fig. 11.** The variation of relative intensity, $I_2$ of trapped positrons as measured from PAL and fraction of Cr atoms surrounding the trapping site, $I_{Cr}$ obtained from CDB ratio curves. The corresponding insets represent data for initial aging period up to 25 h in order to highlight the variations at initial aging times.

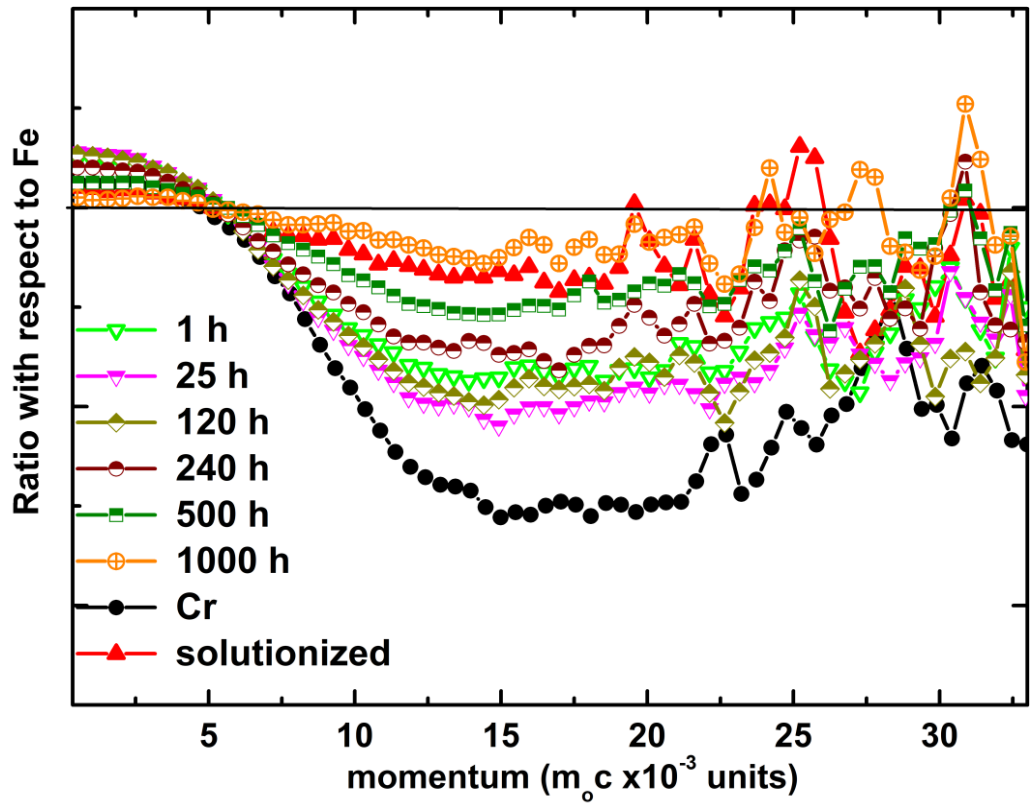

**Fig. 12.** CDB ratio curves (normalized to that of pure defect-free Fe) for Fe-20 at.% Cr alloy aged at 773 K for different durations. The ratio curves of pure Cr and solutionized specimen are also shown for comparison.

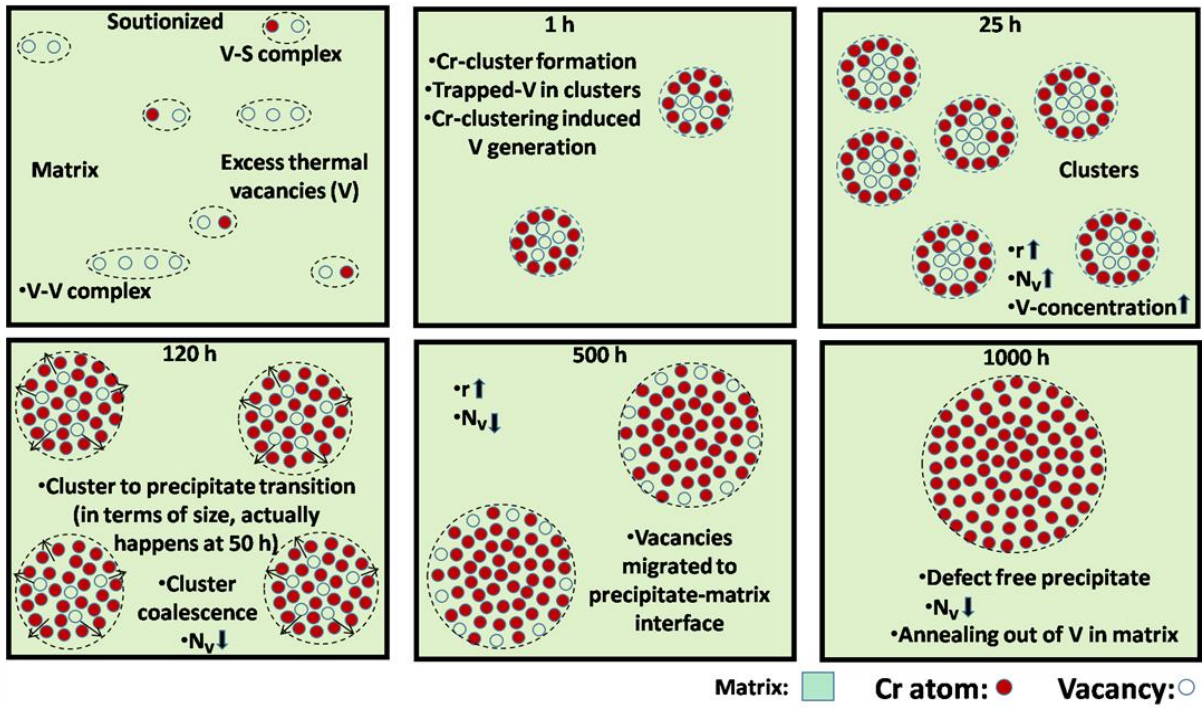

**Fig. 13.** Schematics of vacancy dynamics during the process of non-classical nucleation-growth of α′ in thermally aged Fe-20 at.% Cr alloy.

**Graphical Abstract**

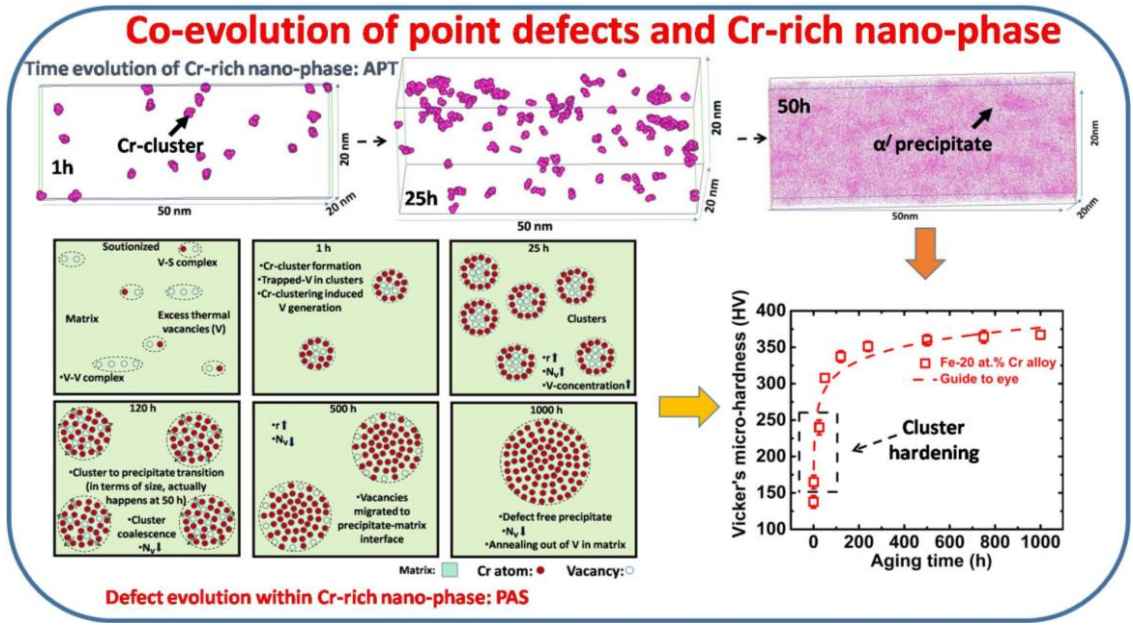